\documentclass[preprint]{ptptex}
\usepackage{graphicx}
\def\del{\partial}

\def\vev#1{\langle #1 \rangle}
\def\del{\partial}

\def\vev#1{\langle #1 \rangle}
\def\del{\partial}
\def\dslash{\del\kern-0.55em\raise 0.14ex\hbox{/}}

\def\rough#1{\raise.3ex\hbox{$#1$\kern-.75em\lower1ex\hbox{$\sim$}}}
\newcommand{\MPL}[3]{{\it Mod. Phys. Lett.} {\bf A{#1}} (19{#3}) {#2}}
\newcommand{\MPLM}[3]{{\it Mod. Phys. Lett.} {\bf A{#1}} (20{#3}) {#2}}
\preprintnumber[3cm]{EHIME-TH-7}
\title{%        %You can use \\ for explicit line-break.
Symmetry and $Z_2$-Orbifolding Approach in 
Five-dimensional Lattice Gauge Theory
}

\author{%       %Use \scshape for the family name.
Kouhei \textsc{Ishiyama}$^{a}$,~%
Michika \textsc{Murata}$^{b}$,~%
Hiroto \textsc{So}$^{c}$
\\and~%
Kazunori \textsc{Takenaga}$^{d}$%
}

\inst{%     %Affiliation, neglected when [addenda] or [errata].
$^a${\it Graduate School of Science and Technology, Niigata University, 
Ikarashi  2-no-cho 8050, Nishi-ku, Niigata 950-2181, Japan.}\\

$^b${\it Support Office for Female Researchers Niigata 
University, Ikarashi  2-no-cho 8050, Nishi-ku, Niigata 950-2181, Japan.}\\

$^c${\it  Department of Physics,  Ehime University, Bunkyou-chou 2-5, 
 Matsuyama 790-8577, Japan.}\\

$^d${\it  Kumamoto Health Science
University, Izumi-machi, Kumamoto 861-5598, Japan.}\\
}

\abst{%         %This abstract is neglected when [addenda] or [errata].
In the framework of 
a pure $SU(2)$ lattice gauge-Higgs unification scenario, 
we find a new global symmetry on a $Z_2$-orbifolded 
five-dimensional space. The global symmetry is consistently 
realized with the $Z_2$-orbifolding, independent of the  bulk gauge symmetry.  
It is shown that the vacuum expectation value of a $Z_2$-projected 
Polyakov loop is a good order parameter for the new symmetry.    
The effective theory on lattice is also discussed.  
}

\begin{document}

\maketitle

\section{Introduction}
The standard model has made a great success and its prediction is 
consistent with all the precision electroweak measurements. The 
model is, however, considered to have potential 
shortcomings, which is related with the Higgs 
sector.  Namely, the Higgs mass  
suffers from ultraviolet effects due to the quadratic 
dependence on the cutoff. The two enormously 
separated energy scales cannot coexist naturally. 
That is the gauge hierarchy problem in the standard model.

Higher dimensional gauge theories have been paid much attention 
as a new approach to overcome the problem without introducing 
supersymmetry. In particular, the gauge-Higgs 
unification \cite{manton, fair, hosotani, gaugehiggs1} is a very 
attractive idea. In the idea, the higher dimensional 
gauge symmetry plays a role to suppress 
the ultraviolet effect on the Higgs mass. 
Moreover, the Higgs self coupling is understood as a 
part of the original higher dimensional 
gauge interaction, so that the mass and the coupling 
can be predicted in the scheme. The gauge-Higgs 
unification has been studied extensively 
from various points of view \cite{gaugehiggs2,gaugehiggs-RS,models}. 

In the scheme, the Higgs field corresponds to the 
Wilson line phase, which is a nonlocal quantity. The Higgs potential is 
generated at the one-loop level after the compactification. Because of 
the nonlocality, the Higgs potential never suffers from the ultraviolet 
effect \cite{masiero}, which is the genuine local effect, and it 
is believed that the Higgs mass calculated from the 
potential is finite as well. In other words, the Higgs mass and the 
potential are calculable in the gauge-Higgs unification. This 
is a remarkable feature which rarely happens in the usual quantum field 
theory. It is understood that the feature entirely comes from shift 
symmetry manifest through the Wilson line phase, which is 
a remnant of the higher dimensional gauge symmetry appeared 
in four dimensions. The Higgs mass does not depend on the cutoff 
at all, so that two tremendously separated energy scales can be 
stable in the gauge-Higgs unification.

The aforementioned attractive property in the 
gauge-Higgs unification is believed to hold in 
perturbation theory\footnote{The finiteness of the Higgs mass 
and potential has been proved at the two-loop level 
in five-dimensional QED with massless fermions \cite{hmty}.}. 
It is natural to ask whether nonperturbative effects destroy the
attractive feature or not. And we are also interested in 
genuine nonperturbative (and/or strong coupling) effects 
on the Higgs mass and the potential\footnote
{In fact, two-loop contributions to the effective potential 
start from the square of the gauge coupling 
constant \cite{hmty}.}. Lattice approach to quantum 
field theories is one of the powerful tools 
to investigate theories nonperturbatively. If we construct an 
effective theory on lattice, we can read off low-energy modes and 
can understand the residual gauge symmetry and the relevant 
particle masses, including gauge and Higgs bosons. 
We believe that nonperturbative studies based on lattice approach 
of the gauge-Higgs unification 
shed some lights on important aspects such 
as finiteness of the Higgs mass, potential and the gauge 
symmetry breaking patterns.

The pioneering works of the lattice approach to the gauge-Higgs 
unification have been done by Irges 
and Knechtli\cite{Irges-Knechtli-1,Irges-Knechtli-2,Irges-Knechtli-3}. 
But they are insufficient to consider the global 
symmetry related to the link variable for the fifth direction 
and the symmetry breaking. One must care a 
relation to the famous Elitzur's theorem\cite{Elitzur} and the gauge 
symmetry  breaking on lattice. The theorem states that  
continuum picture and lattice one are much different 
from each other on gauge fields. 

In this article, a new symmetry in lattice gauge theories 
with $Z_2$-orbifolding is presented. It is a discrete 
and global symmetry, independently of the gauge symmetry. 
Owing to the new symmetry, the associated theorem on physical quantities 
such as correlation functions of a Polyakov loop are proved. In 
the next section, we present the lattice version of 
a five-dimensional $SU(2)$ gauge-Higgs unification with an orbifold 
compactification $S^1/Z_2$, paying attention to the global 
symmetry which is essential in our lattice approach. 
We find a new symmetry and present a theorem 
led from the new symmetry in section $3$. In section $4$ we discuss 
an effective lattice theory using the new symmetry. 
In doing it, the Elitzur's theorem comes 
into play. The final section is devoted to summary and discussions. 
%%%%%%%%%%%%
\section{Formulation}
\subsection{Orbifolding on lattice}
Let us present the lattice formulation of the $SU(2)$ gauge-Higgs
unification compactified on the orbifold $S^1/Z_2$ in this section. 
The $S^1$ topology imposes a periodic boundary condition 
on a lattice field 
\begin{equation}\label{eq:PBC} 
\Phi_{n_M} = \Phi_{n_M+{N_5\hat{5}}} ~, 
\end{equation}
where lattice coordinates and the lattice size for the
fifth direction are written as $n_M = 
\{n_{\mu}, n_5\}$ and $N_5$, respectively. We also use a 
notation $M=(\mu,5)$ for directions and set the lattice 
constant $a$ unity. Here we consider our lattice model 
as a cutoff theory according to  Irges and 
Knechtli\cite{Irges-Knechtli-1,Irges-Knechtli-2,Irges-Knechtli-3}.
The $S^1/Z_2$ compactification 
is implemented by a reflection operator $\cal{R}$ and a group 
conjugation operator 
${\cal{T}}_{g_0}$ 

\begin{equation}\label{eq:PRJ} 
\frac{1-\Gamma}{2}U_{n_M,N}=0,~~\Gamma \equiv {\cal{R}}{\cal{T}}_{g_0} ~. 
\end{equation}
\noindent
In order to insure $\Gamma^2 = 1$, the operators should satisfy 
$[{\cal{R}}, {\cal{T}}_{g_0} ]=0$ and ${\cal{R}}^2 = {\cal{T}}^2_{g_0} = 1$. 
The reflection operator acts for the coordinate as 
\begin{equation}\label{eq:rfl}
{\cal{R}} n_M = \bar{n}_M\equiv \{n_{\mu}, -n_5\} ~.
\end{equation}
\noindent 
Taking accounting of the periodicity by $N_5$ for the fifth 
coordinate, we find two fixed points, $n_5 = 0$ 
and $n_5 = N_5/2 \equiv  L_5 $ whose four-dimensional subspaces 
are invariant under $\cal{R}$. For link variables, $\cal{R}$ acts as 
\begin{eqnarray}\label{eq:Reflection}
{\cal{R}}U_{n_M,\nu}&=& U_{\bar{n}_M,\nu}  ~, \nonumber\\
{\cal{R}}U_{n_M,5} &=& U^{\dagger}_{\bar{n}_M-\hat{5},5}~, \nonumber\\
{\cal{R}}U^{\dagger}_{n_M,5} &=& U_{\bar{n}_M-\hat{5},5}~,
\end{eqnarray}
\noindent 
and the group conjugation operator ${\cal T}_{g_0}$ acts as
\begin{equation}\label{eq:GPRJ} 
{\cal T}_{g_0}U_{n_M,N} = g_0U_{n_M,N}g_0^{\dagger}  ~.
\end{equation}
Here $g_0^2$ must be an element of center 
group in $SU(2)$ by the condition ${\cal{T}}^2_{g_0} = 1$~.

A nontrivial choice $g_0 = i\sigma_3$ induces a breaking 
of $SU(2)$ symmetry to $U(1)$ symmetry at two fixed 
points $n_5 = 0$ and $L_5$, which are called as $FP(1)$ and $FP(2)$, 
respectively. This is a typical symmetry breaking mechanism 
by orbifolding.\cite{orbi} 
By this $S^1/Z_2$ orbifold compactification, the starting 
action with $S^1$ compactification in five dimensions 
\begin{equation}\label{eq:S1action} 
S_{S^1} =\beta\sum_{P\in S^1 }[1-\frac{1}{2}{\rm~ Tr~}U_P]  
\end{equation}
\noindent
becomes
\begin{equation}\label{eq:S1/Z2action}
S_{S^1/Z_2} = \beta\sum_{P\in {\rm bulk~in~} S^1/Z_2}[1-\frac{1}{2}{\rm~ Tr~}U_P] 
+ \frac{\beta}{2}\sum_{P\in FP(1),FP(2)}[1-\frac{1}{2}{\rm~ Tr~}U_P] ~, 
\end{equation}
\noindent
where the $U_P$ implies the product of link variables for 
a plaquette $P$. For the link variable $U_{{n_{\mu}},\nu}(I)$ on 
each fixed point $FP(I)(I=1, 2)$,  it is reminded of the condition 
\begin{equation}\label{eq:FPlinkcondition}
U_{n_{\mu},\nu} (I) = g_0U_{n_{\mu},\nu}(I)g_0^{\dagger}~,
\end{equation}
\noindent
which is followed from the $Z_2$-projection (\ref{eq:PRJ}). The 
condition (\ref{eq:FPlinkcondition})  
restricts $U_{n_{\mu},\nu} (I) $  to $U(1)$-values.  
The link variable is locally transformed under the $U(1)$ as 
\begin{equation}\label{eq:U1trans}
U'_{n_{\mu},\nu} (I) = u(n_{\mu},I)U_{n_{\mu},\nu} 
(I) u^{\dagger}(n_{\mu}+\hat{\nu},I),~~I=1,2~.
\end{equation}
\noindent
Here $u(n_{\mu},I)$ is an $U(1)$ element that depends 
on a four-dimensional coordinate $n_{\mu}$ 
and $[g_0, u(n_{\mu},I)] = 0$. It is easy to see
that (\ref{eq:U1trans}) keeps the action (\ref{eq:S1/Z2action}) 
invariant and is consistent with (\ref{eq:FPlinkcondition}).   
One can verify that the action (\ref{eq:S1/Z2action}) is invariant
under a remained bulk $SU(2)$ gauge symmetry    
\begin{eqnarray}\label{eq:SU2trans}
U'_{n_M,N}=\left\{
\begin{array}{lll}
 V_{n_M}U_{{n_M},\nu}V^{\dagger}_{{n_M}+\hat{\nu}} & {\rm ~for~} 
N=\nu {\rm ~and~} n_5\ne 0,L_5~, &\quad  \\
 U_{{n_M},\nu} &{\rm ~for~} N=\nu {\rm ~and~}      n_5=0,L_5 ~,&\quad  \\
  V_{n_M}U_{{n_M},5}V^{\dagger}_{{n_M}+\hat{5}}  &{\rm ~for~}
N=5 {\rm ~and~} n_5\ne -1,0,L_5-1,L_5~,&\quad  \\
 U_{{n_M},5}V^{\dagger}_{{n_M}+\hat{5}} &{\rm ~for~}N=5 {\rm ~and~} 
n_5=0,L_5~,&\quad  \\
  V_{n_M}U_{{n_M},5} &{\rm ~for~}N=5 {\rm ~and~} n_5=-1,L_5-1~.&\quad
\end{array}
\right.
\end{eqnarray}
\noindent
%%%%%%%%%%%%%%%%%%%%%%%%%%%
\subsection{Order parameter of our model}
The compactness of the five-dimension apparently 
indicates that a Polyakov loop
\begin{equation}\label{eq:PL}
L(n_{\mu})\equiv {\rm ~Tr~}U_{\{n_{\mu},0\},5}\cdots U_{\{n_{\mu},2L_5-1\},5}
\end{equation}
\noindent
is an order parameter for the center symmetry defined by 
%
%for an arbitrary but fixed $n_5 = k$
%
\begin{equation}\label{eq:center-trans}
U'_{{n_M},N}=\left\{
\begin{array}{ll}
U_{{n_M},\nu}  & {\rm ~for~}N=\nu ~,  \quad  \\
zU_{{n_M},5}  &   {\rm ~for~}N=5,~ n_5=k~, \quad  \\
U_{{n_M},5}    &  {\rm ~for~}N=5,~ n_5\ne k~,\quad
\end{array}
\right.
\end{equation}
\noindent
where an element $z$ is the center group. We must 
take into account of the $Z_2$-projection 
(\ref{eq:PRJ}) in $(11)$ for the case of the orbifold 
$S^1/Z_2$. Then the loop is rewritten as 
%%%%%
%%%%%%%%%
\begin{equation}\label{eq:PL-Z2}
L_2(n_{\mu}) \equiv {\rm ~Tr~}U_{\{n_{\mu},0\},5}\cdots 
U_{\{n_{\mu},L_5-1\},5}g_0 U^{\dagger}_{\{n_{\mu},L_5-1\},5}
\cdots U^{\dagger}_{\{n_{\mu},0\},5}g_0^{\dagger}  ~.
\end{equation}
\noindent
This expression (\ref{eq:PL-Z2})  is called as 
a $Z_2$-projected Polyakov loop. Contrary to the Polyakov loop $(11)$,
the $Z_2$-projected Polyakov loop is invariant 
under (\ref{eq:center-trans}) because it always has a pair of $U_{{n_M},5}$ 
and $U^{\dagger}_{{n_M},5}$ with $n_5=k$. Hence the 
loop (\ref{eq:PL-Z2}) is not suitable for an order parameter 
of the center symmetry.
%
%%%%%%%%%%%
\begin{figure}
\begin{center}
\vspace{0.5cm}
\includegraphics[width=8cm]{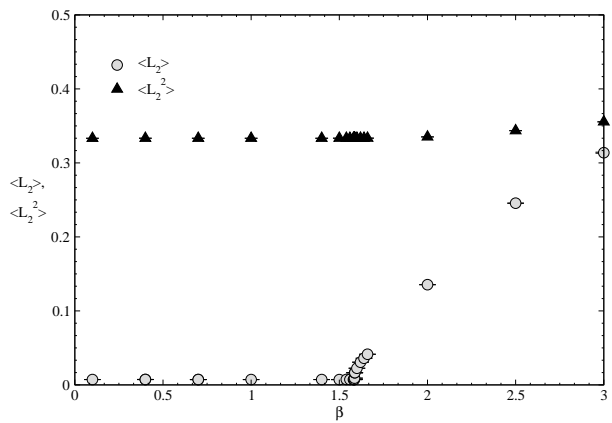}
\caption{
Vacuum expectation values of $L_2$ (gray circles) 
and $L_2^2$ (black triangles) at $8^4  \times 8 (L_5=4)$  
lattice.}
\end{center}
\end{figure}
%\noindent

The $Z_2$-projected Polyakov loop (\ref{eq:PL-Z2}) and its square
have been computed on a $8^4 \times 8$ lattice (namely, $L_5=4$) by using 
Monte-Carlo simulation with heatbath and overrelaxation 
algorithms~(Fig.~1). Clearly, for $\beta < \beta_c \approx 1.6$, the 
vacuum expectation value (VEV) of the loop is vanishing and 
for $\beta > \beta_c$, it is increasing. $\beta_c$ is considered 
as a critical coupling. It is noted that 
not only  the VEV of the loop but also that of square 
are very stable for $\beta < \beta_c$, whose coupling region 
implies the confining phase.

\section{New symmetry and stick theorem}
The argument of the previous section apparently leads 
to a conclusion that the $Z_2$-projected Polyakov loop is 
unsuitable for the order parameter of the center 
symmetry (\ref{eq:center-trans}). The result of the 
Monte-Carlo simulation~(Fig.~1), however, implies that 
there  is  {\it a certain symmetry}. We 
shall clarify the explicit form and the property of the 
symmetry in this section. 

\subsection{Stick transformation and new symmetry}

At first, an expected transformation for link 
variables is independent of (\ref{eq:SU2trans}).
The explicit form is  
\begin{equation}
\label{eq:newsymmetry}
U'_{{n_M},N}=
\left\{
\begin{array}{ll}
\alpha(n_{\mu},I)U_{{n_{\mu}},\nu}(I)\alpha^{\dagger}(n_{\mu}+\hat{\nu},I)& 
{\rm for~links~ on~}FP(I)~,I=1,2~, 
\\
\alpha(n_{\mu},1)U_{n_M,5} &   {\rm for~sticking~links~ out~}FP(1)~,   
\\
U_{{n_M},5}\alpha^{\dagger}(n_{\mu},2) & {\rm for~sticking~links~into~}FP(2)~,
\\
U_{{n_M},\nu} & {\rm for}~n_5=1,\cdots,L_5-1~{\rm with}~N=\nu~,\\
U_{{n_M},5} &{\rm  for}~n_5=1,\cdots,L_5-2~{\rm with}~N=5~,
\end{array}\right.
\end{equation}
where $\alpha(n_{\mu}, I)$ is an element of the $SU(2)$.
Let us note that the above transformations are 
defined for the links on the orbifold $S^1/Z_2$, not on the
$S^1$. The transformations for other links are
determined by the $Z_2$ projection (\ref{eq:PRJ}).

%%%%%%
%%%%%%%%%
The first transformation of (\ref{eq:newsymmetry}) must be 
careful in the consistency with (\ref{eq:FPlinkcondition}), 
%%%%%%%%
\begin{equation}\label{eq:consistency}
g_0\alpha(n_{\mu},I)=\alpha(n_{\mu},I)g_0z(n_{\mu},I) ~.
\end{equation}
\noindent
Here $z(n_{\mu},I)$ is an element of $SU(2)$ which commutes with 
any $U_{{n_{\mu}},\nu}(I)$ \footnote{In a group theoretical 
terminology, $z(n_{\mu},I)$ belongs to a {\it centralizer} 
with $U(1)$, $i.e.$ $[z(n_{\mu},I),U_{{n_{\mu}},\nu}(I)]=0$.}.

The action (\ref{eq:S1/Z2action}) is invariant under (\ref{eq:newsymmetry}).
This is because the plaquettes on the fixed points are invariant under the 
first transformation in (\ref{eq:newsymmetry}) and the plaquette
oriented for the fifth direction from the $FP(2)$
\begin{eqnarray}
  U_P
  =U_{\{n_\mu,\,L_5-1\},\,5}
   U_{n_\mu,\,\nu}(2)
   U^\dagger_{\{n_\mu+\hat{\nu},\,L_5-1\},\,5}
   U^\dagger_{\{n_\mu,\,L_5-1\},\,\nu}
\end{eqnarray}
is also invariant under the first and the third transformations
\begin{eqnarray}
U_P\rightarrow U_P'&=&
    U_{\{n_\mu,\,L_5-1\},\,5}\alpha^{\dagger}(n_{\mu},2)
    \alpha(n_{\mu},2) U_{n_\mu,\,\nu}(2)
    \alpha^{\dagger}(n_{\mu}+\hat{\nu},\,2)
    \nonumber\\ &\mbox{}&\qquad\times
    \alpha(n_{\mu}+\hat{\nu},\,2) 
    U^\dagger_{\{n_\mu+\hat{\nu},\,L_5-1\},\,5}
    U^\dagger_{\{n_\mu,\,L_5-1\},\,\nu}\,.
\end{eqnarray}
%%%%%%%%%
%%%%%%
Hereafter we use the terminology, the FP gauge symmetry 
instead of the $U(1)$ gauge symmetry. 
It is important to note that the first 
transformation in (\ref{eq:newsymmetry}) pulls the 
$U_{{n_{\mu}},\nu}(I)$ back to $U(1)$ of the FP 
gauge symmetry. 

The explicit solutions for 
(\ref{eq:consistency}) are 
\begin{equation}\label{eq:solution}
\alpha(n_{\mu},I)=
\left\{
\begin{array}{ll}
e^{i\theta(n_{\mu},I)\sigma_3} &  {\rm ~for~}z(I)=1{\rm ~case} ~,  
\quad  \\
(i\sigma_2)e^{i\theta(n_{\mu},I)\sigma_3} &   
{\rm ~for~}z(I)=-1{\rm ~case}~, \quad \\
{\rm no~solution~}    &  {\rm ~for~other~cases}  ~. \quad
\end{array}
\right.
\end{equation}
The first case in (\ref{eq:solution}) 
just corresponds to the $U(1)$ gauge transformation
(\ref{eq:U1trans}). The second case is essentially a 
new global symmetry (up to the $U(1)$ gauge transformation). 
The consistency between (\ref{eq:FPlinkcondition}) 
and (\ref{eq:newsymmetry}) is confirmed as 
\begin{eqnarray}\label{eq:FPconsistent}
g_0U'_{n_{\mu},\nu}(I)g_0^{\dagger}  
&= & g_0\alpha(n_{\mu},I)  U_{n_{\mu},\nu}(I) 
\alpha^{\dagger}(n_{\mu}+\hat{\nu},I)g_0^{\dagger}  \nonumber   \\
&=& \alpha(n_{\mu},I) g_0z(n_{\mu},I) 
U_{n_{\mu},\nu}(I)z^{\dagger}(n_{\mu}+\hat{\nu},I) g_0^{\dagger} 
\alpha^{\dagger}(n_{\mu}+\hat{\nu},I)  \nonumber \\
&=&  U'_{n_{\mu},\nu}(I) ~,
\end{eqnarray}
\noindent
where $[z(I),U_{n_{\mu},\nu}(I)]=[g_0,U_{n_{\mu},\nu}(I)]=0$ 
has been used. This symmetry is global not local because the 
last equality in (\ref{eq:FPconsistent}) holds only under the 
condition
\begin{equation}\label{eq:z-consistency}
z(n_{\mu},I)z^{\dagger}(n_{\mu}+\hat{\nu},I)=1~. 
\end{equation}
\noindent
In order to obtain the nontrivial transformation for the  
$Z_2$-projected Polyakov loop $L_2(n_{\mu})$, we 
adopt $z(1)\ne z(2)$ as we will see below. 
Under our assignment of  $z(1)=1$ and $z(2)=-1$,  the 
transformation (\ref{eq:newsymmetry}) 
(up to the $U(1)$ gauge transformation) becomes 
\begin{equation}\label{eq:globalnewsymmetry}
U'_{{n_M},N}=
\left\{
\begin{array}{ll}
(i\sigma_2)U_{n_{\mu},\nu}(2)(-i\sigma_2) & {\rm for~links~on~}FP(2)~, \\
U_{\{n_{\mu},0\},5} &
{\rm  for~sticking~ links ~out~}FP(1)~,  \\
U_{\{n_{\mu},L_5-1\},5}(-i\sigma_2) &
{\rm  for~sticking~ links ~into~}FP(2)~,  \\
U_{{n_M},\nu} & {\rm for}~n_5=0,\cdots,L_5-1~{\rm with}~N=\nu~,\\
U_{{n_M},5} &{\rm  for}~n_5=1,\cdots,L_5-2~{\rm with}~N=5~.
\end{array}
\right.
\end{equation}
\noindent
Here link variables $U_{n_{\mu},\nu}(1)$  are transformed as the fourth  case. 
This global transformation   (\ref{eq:globalnewsymmetry}) is called as {\it stick } one, where  
we have defined a discrete transformation {\it \'{a} la} 
stick \footnote{The counterpart of the discrete 
transformation seems to be unknown in the continuum theory.}.
Another assignment $z(1)=-1, z(2)=1$ is equivalent 
to (\ref{eq:globalnewsymmetry}) after the change of 
variable generating the exchange, $z(1)\leftrightarrow z(2)$.

The action (\ref{eq:S1/Z2action}) has four 
symmetries (\ref{eq:U1trans}), (\ref{eq:SU2trans}), (\ref{eq:center-trans})
and the new global symmetry (\ref{eq:globalnewsymmetry})
that we call the stick symmetry.
%%%%%%%%%
%%%%%%%%%%
With respected to the path-integral 
measure $dU_{{n_{\mu}},\nu}(2)$ for the stick transformation,     
we can understand  the invariance from a fact 
that (\ref{eq:globalnewsymmetry}) induces an isomorphic 
map from a compact $U(1)$ into another compact $U(1)$  
for link variables on $FP(2)$\footnote{The 
invariance of the measure is clear because  
a stick transformation of link variables on $FP(2)$ 
by (\ref{eq:globalnewsymmetry}) is equivalent 
to $\theta(n_{\mu},\nu) 
\leftrightarrow -\theta(n_{\mu},\nu)$, where 
$U_{{n_{\mu}},\nu}(2)=e^{i\theta(n_{\mu},\nu)\sigma_3}$.}.

The $Z_2$-projected Polyakov loop is transformed nontrivially 
under (\ref{eq:globalnewsymmetry}) as
\begin{eqnarray}\label{eq:PLnewtrans}
%\lefteqn{} \\
L'_2(n_{\mu})&=& {\rm Tr~}U_{\{n_{\mu},0\},5}\cdots 
U_{\{n_{\mu},L_5-1\},5}(-i\sigma_2)g_0(i\sigma_2) 
U^{\dagger}_{\{n_{\mu},L_5-1\},5} \cdots 
U^{\dagger}_{\{n_{\mu},0\},5}g_0^{\dagger} \nonumber\\ 
&=&-L_2(n_{\mu}).
\end{eqnarray} 
\noindent 
This means that the loop can be an order 
parameter for the stick symmetry. 
At first glance, the stick symmetry seems to be a subgroup 
of the bulk gauge symmetry, but it is never a gauge symmetry
and is actually an independent global symmetry, as shown 
by (\ref{eq:globalnewsymmetry}) and  (\ref{eq:PLnewtrans}). 
%%%%
%%%%%%%
Here let us summarize the symmetry properties of the Polyakov 
loop in $S^1, S^1/Z_2$ models under the center 
and stick transformations in Table I.
%%%%%%%
%%%%%%%%%%
%\begin{center}
\begin{table}[htbp]
\begin{center}
\begin{tabular}{|c|c|c|c|} \hline
models & Polyakov loop & center symmetry  &  new (stick) symmetry \\ \hline
$S^1$ &Tr~$UUUU\cdots$ & variant  &  not defined   \\ \hline
$S^1/Z_2$ &  Tr~$UU\cdots g_0U^{\dagger}U^{\dagger}\cdots 
g_0^{\dagger}$& invariant  & variant \\ \hline
\end{tabular}
\caption{Comparison between $S^1$ model and $S^1/Z_2$ model}
\end{center}
\end{table}
%%%%%%%%%%%%%%%%%
%%%%%%%%

\subsection{Stick theorem and sticking operators}

Correlation functions between two  $Z_2$-projected Polyakov 
loops are important quantities since they may be related to 
Higgs fields and their masses. The new symmetry
(\ref{eq:globalnewsymmetry}) controls not 
only the VEV of a single $Z_2$-projected Polyakov loop, but 
also the VEVs of the correlation function of the loops.  
%%%%%%%%%%%%
%
%%%%%%%%%%%%%%%%
The fundamental property of the VEVs of sticking operators 
into the $FP(2)$ (Fig. 2) is stated as a stick theorem:

\vspace{0.5 cm}
{\it A VEV of any product operators made from link variables 
sticking into the $FP(2)$ odd number of times vanishes unless the stick 
symmetry (\ref{eq:globalnewsymmetry}) is broken.}

\vspace{0.5 cm}
%%%%%%%%%%%%%%%%%%%%%
%%%%%%%%%%%%%%%%%%%%%%%
\noindent
In order to prove this theorem, let us consider any operator $F(U)$
consisted of the link variables sticking into the $FP(2)$ odd number
of times $N_o$
\begin{eqnarray}\label{eq:oddPL}
F(U) \equiv & ({\rm Tr~}M_1 U_{\{n_{\mu},L_5-1\},5}g_0 
U^{\dagger}_{\{n_{\mu},L_5-1\},5}M_1^{\dagger}g_0^{\dagger} )
\cdots \nonumber\\
& ({\rm Tr~}M_2 U_{\{n'_{\mu},L_5-1\},5}g_0 
U^{\dagger}_{\{n'_{\mu},L_5-1\},5}M_2^{\dagger}g_0^{\dagger})~,
\end{eqnarray}
\noindent
where $M_1$ and $M_2$ mean various product operators made from  
link variables detached from the $FP(2)$.
%%%%%%%%%%% 
%%%%%%%%%%%%
We find by executing the change of variables
with (\ref{eq:globalnewsymmetry}) that 
\begin{eqnarray}\label{eq:odd-trans} 
<F(U)>&=&<F(U')> \nonumber \\
&=&\int \prod\limits_{n_M ,N}  dU'_{n_M,N} e^{-S_{S^1/Z_2}(U')} F(U')
\left/   
\int \prod\limits_{n_M ,N}  
dU'_{n_M,N} e^{-S_{S^1/Z_2}(U')}\right. \nonumber \\
&=& (-1)^{N_o}\int \prod\limits_{n_M ,N}  dU_{n_M,N} e^{-S_{S^1/Z_2}(U)} F(U)
\left/ 
\int \prod\limits_{n_M ,N}  dU_{n_M,N} e^{-S_{S^1/Z_2}(U)} 
\right. \nonumber \\
&=&  - <F(U)>~, 
\end{eqnarray}
\noindent
where $S_{S^1/Z_2}(U)$ is an invariant plaquette action 
under (\ref{eq:globalnewsymmetry}). An equation (\ref{eq:odd-trans}) means
that the VEV of the operator (\ref{eq:oddPL}) vanishes 
if the stick symmetry (\ref{eq:globalnewsymmetry}) is unbroken. 
\noindent
\vspace{0.3 cm}
\begin{flushright}
q.e.d.
\end{flushright}
%\noindent
% 
%
\begin{figure}[htbp]
\begin{center}
\includegraphics[width=7.0cm]{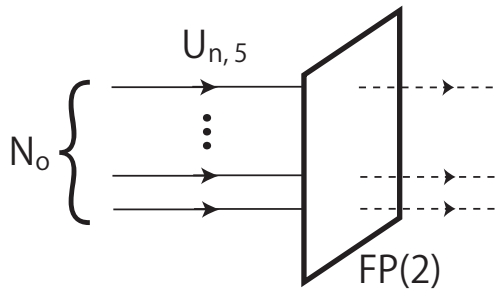}
\caption{The $Z_2$-projected Polyakov loops 
sticking odd number of times into the fixed point $FP(2)$.
The dashed arrows are related with $U_{n, 5}$ by the 
$Z_2$ projection (\ref{eq:PRJ}).}
\end{center}
\end{figure}
%%%%%
%\noindent
%%%%%
The simplest example of the sticking operator into the 
$FP(2)$ is the $Z_2$-projected Polyakov loop, which corresponds
to the case with $N_o=1$ in Fig. $2$.

Furthermore, the sticking operators have some important
properties for constructing the effective theory. The operators
are stable against corrections in the strong coupling regime.
The VEV of any product 
operator $F$ made from the link variables sticking 
into the $FP(2)$ locally odd number of times $N_o$ vanishes 
in the strong coupling limit owing to ${\rm ~Tr~}g_0=0$ and
$g_0^2=-1$. Here the {\it locally} ~odd number of times means odd
number of times 
sticking into a four-dimensional point 
on the $FP(2)$. But it admits the link variables to stick 
into the whole $FP(2)$ even number of times (Fig. 3). 
The typical example of 
such the operator is the correlation function 
of the $Z_2$-projected Polyakov loop. In the strong coupling limit, 
it is needless to consider any corrections for the plaquette. 
\begin{figure}[htbp]
\begin{center}
\includegraphics[width=6cm]{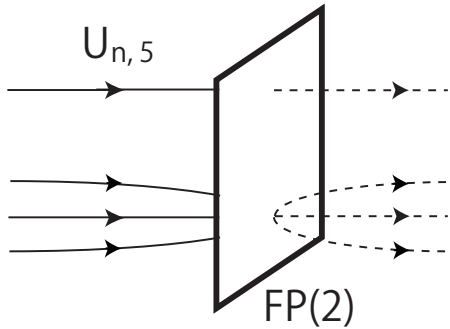}
\caption{The $Z_2$-projected Polyakov 
loops sticking {\it locally} odd number of times into the fixed point 
$FP(2)$. The dashed arrows are related with $U_{n, 5}$ by the 
$Z_2$ projection (\ref{eq:PRJ}).}
\end{center}
\end{figure}
\\
%%%%%%%
%%%%%%
We also observe from Fig. 1 that $<L_2^2>$, which sticks 
{\it locally}~even number of times into the $FP(2)$, seem to be 
very stable for $\beta < \beta_c$~.
%%%%%%%%%
%
%%%%%%%%%%
Before closing this section, it may be meaningful to state 
that the VEVs of the sticking operators into 
the $FP(2)$ always suppress any strong coupling corrections 
in the confining phase. This implies that the sticking 
operators are good candidates to describe the effective
theory in the phase. These results are useful for constructing the
effective theory, which will be discussed in the next section. 

\section{Effective theory and Elitzur's theorem}

Based on the new global symmetry (\ref{eq:globalnewsymmetry}), let us 
construct an effective theory from our lattice model in this section. 
Before proceeding with it, it may be instructive to mention the 
naive continuum (perturbative) limit of (\ref{eq:PRJ}). 

\subsection{Naive continuum limit and an effective theory}

If we write $U_{n_M, N}={\rm exp}(iaA_N(n_M))$
in (\ref{eq:Reflection}) and (\ref{eq:GPRJ}), then we find that 
boundary conditions for the gauge potential $A_N$,
\begin{equation}  \label{eq:Amu-A5}
A_{\nu}(n_{\mu}, n_5)=g_0A_{\nu}(n_{\mu}, -n_5)g_0^{\dagger}~,
\qquad
A_5(n_{\mu}, n_5)=-g_0A_5(n_{\mu}, -n_5)g_0^{\dagger}~.
\end{equation}
For $g_0=i\sigma_3$, the gauge symmetry is broken down to $U(1)$
by the orbifolding \cite{orbi}. The zero modes in $A_5$, which
are actually given by $A_5^1$ and  $A_5^2$ 
from (\ref{eq:Amu-A5}), play the role of the Higgs field in 
the gauge-Higgs unification. 
%%%%%%%
%
%%%%%%%%%%
The Wilson line phase is an important quantity
in the gauge-Higgs unification and can be 
written by the zero mode of the gauge 
potential $A_5$ ({\it \'{a} la} Higgs field). 
It is explicitly obtained 
by \footnote{The $\sigma_3$ part of $A_5$ has no zero mode 
in the continuum theory.} 
\begin{equation}\label{eq:WL}
W_c = {\cal P} \mbox{exp} 
\left(ig_5 \oint_{S^1/Z_2}dy \vev{A_5} \right)
={\cal P} \mbox{exp} 
\left(ig_5 \oint_{S^1/Z_2}dy \biggr(\vev{A_5^1}{\sigma_1\over 2}
+\vev{A_5^2}{\sigma_2\over 2}\biggl) \right)~,
\end{equation}
where $g_5$ is the five-dimensional gauge coupling and 
${\cal P}$ stands for the path-ordered product. We stress that 
the phase $W_c$ does not correspond  to the loop (\ref{eq:PL-Z2}), 
but to an operator $X(n_{\mu})$ defined below. 
As the result, the effective theory in the naive 
continuum limit can be expressed by 
$<A_5^1>,~<A_5^2>$ and $A_{\mu}^3$. 
%%%%%%%%%%%%%%%%%%%%%%%%%%%%%%%%%%%
%
%%%%%%%%%%%%%%%%%%%%%%%%%%%%%

\subsection{Elitzur's theorem and its generalization}

Since the notion of the gauge invariance is crucial on lattice, 
the physical picture based on the zero modes or the 
VEVs of gauge fields alone is useless because they are
gauge variant quantities. 
%%%%%%%%%%%%%%%%%%%%%
%
%%%%%%%%%%%%%%
The crucial point on lattice gauge theories is the existence of  
a theorem by Elitzur\cite{Elitzur} on the VEV of a single link variable 
on lattice. The theorem precisely 
states that the VEV of the variable vanishes 
whenever the local symmetry is kept 
on lattice\footnote{There is no counterpart of the theorem
in the continuum theory
%%%%%%
%
%%%%%%%
because it is difficult to control both ultraviolet and infrared 
divergences simultaneously.}. For a composite operator 
made from link variables such as 
$U_{\{n_{\mu},L_5-1\},5}g_0U^{\dagger}_{\{n_{\mu},L_5-1\},5}g_0^\dagger$, 
a similar theorem holds except for the singlet component 
after the decomposition of the operator into 
irreducible representations, {\it i.e.},
the VEV of the nontrivial components vanishes 
whenever the local symmetry is kept on lattice.

In our case, we need to generalize the Elitzur's theorem 
to  a five-dimensional lattice gauge theory 
with the FP  gauge symmetries in the four-dimensional lattice spaces. 
%%%%%%%
% 
%%%%%%%%%%
A generalized statement on the Elitzur's theorem follows as:  

\vspace{0.3 cm}
{\it When we consider the lattice gauge transformation 
by the subgroup of an original gauge group,   
the VEVs of nontrivially gauge transformed operators  are vanishing. }

\vspace{0.5 cm}
\noindent
The proof of this generalization is essentially the 
same as  Elitzur's original one except for 
the consideration of  the lattice gauge transformation 
corresponding to  the subgroup. 

%%%%%%%%%%
%\par \noindent
%%%%%%%%%
From this generalization,  we can understand 
that any FP gauge symmetry is always unbroken in the lattice gauge theory 
with the $Z_2$-orbifolding, because our FP gauge symmetry can be regarded as 
a subgroup of a bulk  gauge transformation. 
The global stick symmetry is independent of 
the bulk gauge transformation and possible to be broken spontaneously.

\subsection{Lattice effective theory}

The effective theory must be constructed by 
gauge invariant operators such as a trace of 
the plaquette $U_P$ and a $Z_2$-projected 
Polyakov loop and by low-energy modes.
%%%%%%
%
%%%%%%%%%
Not only in a confining phase but also in a deconfining 
phase, these 'effective' operators must be gauge invariant and/or 
must be the constituent parts of low-energy effective action with 
the gauge invariance on lattice. On the other 
hand, the zero modes of the component gauge fields for the fifth
direction
are important on the continuum theory. However, the zero 
mode is not gauge invariant, so that it cannot
be consisted of a part of the low-energy effective action with the
gauge invariance on lattice. Instead of the 
zero mode, we define an operator  
\begin{equation}\label{eq:PL-1/2}
X(n_{\mu}) \equiv U_{\{n_\mu  ,0\},5} U_{\{n_\mu  ,1\},5}  
\cdots U_{\{n_\mu, L_5 - 2\},5} U_{\{n_\mu  ,L_5 - 1\},5}  ~.
\end{equation}
\noindent
It is noted that the $X(n_{\mu})$ is a bi-fundamental field 
for the FP gauge symmetry 
\begin{equation}\label{eq:bi-fundamentalforFP}
X'(n_{\mu})=e^{i\theta(n_{\mu},1)\sigma_3}X(n_{\mu})
e^{-i\theta(n_{\mu},2)\sigma_3} ~,
\end{equation}
\noindent
and is transformed as 
\begin{equation}\label{eq:X field sticksymmetry}
X'(n_{\mu})=X(n_{\mu})(-i\sigma_2)~, 
\end{equation}
\noindent
under the stick transformation. From (\ref{eq:PL-1/2}), we can express the
$Z_2$-projected Polyakov loop (\ref{eq:PL-Z2}) as 
\begin{equation}\label{eq:PLby1/2}
L_2(n_{\mu}) = {\rm Tr~}  X(n_{\mu})g_0X^{\dagger}(n_{\mu})g_0^{\dagger}~, 
\end{equation}
\noindent
which is clearly the FP gauge invariant and odd for the stick 
symmetry. From the discussion of the previous section including 
the stick theorem, the loop $L_2(n_{\mu})$ and the operator $X(n_{\mu})$  
are very stable in the confining phase. And the effective
potential for the loop should be an even function for the stick symmetry. For  
the pure $FP(I)$ gauge sector, the simplest FP gauge and 
stick symmetry invariant operators are traces of plaquette 
\begin{equation}\label{eq:plaquette}
{\rm Tr~}U_P(I) \equiv 
{\rm Tr~}U_{n_{\mu},\nu}(I)
U_{n_{\mu}+\hat{\nu},\rho}(I)
U_{n_{\mu}+\hat{\rho},\nu}^{\dagger}(I)
U_{n_{\mu},\rho}^{\dagger}(I) ~{\rm ~for~}I=1,2~,
\end{equation}
\noindent
where the stick symmetry implies that 
\begin{eqnarray}\label{eq:stick for link}
U'_{n_{\mu},\nu}(1)&=&U_{n_{\mu},\nu}(1)~,  \nonumber \\
U'_{n_{\mu},\nu}(2)&=&(i\sigma_2)U_{n_{\mu},\nu}(2)
(-i\sigma_2)=U^*_{n_{\mu},\nu}(2)  ~.
\end{eqnarray}
\noindent
It is noted that 
the ${\rm Tr~}U_P(2)$ on the $FP(2)$ is real because 
the link variable $U_{n_{\mu},\nu}(2)$ 
belongs to the subgroup $U(1)$ of the $SU(2)$. 

The first stage to construct the effective theory is to find
massless or light modes. The massless modes are massless gauge fields
associated with the FP gauge symmetry. The link variables which are variant 
under the $SU(2)$ bulk gauge symmetry (\ref{eq:SU2trans}) are
path-integrated out.  
%%%%%%
%
%%%%%%%%
We assume that the variable $X(n_{\mu})$ defined by $(26)$, which
is invariant under (\ref{eq:SU2trans}), is a fundamental 
operator in the effective theory.
%%%%% 
%
%%%%%%
The second stage is to look for the form of the couplings among 
the modes. From the action (\ref{eq:S1/Z2action}), the link 
variable $U_{n_{\mu},\nu}(I)$ is coupled with
staple products of the link variables which transform 
as the bi-fundamental representation of the FP gauge
symmetry. Assuming that the staple products can be replaced by 
$X^{\dagger}(n_{\mu}+\hat{\nu})U^{\dagger}_{n_{\mu},\nu}(1)X(n_{\mu})$, 
the effective theory can be written as  
\begin{eqnarray}\label{eq:Seff}
S_{{\rm eff}}&=& \sum_{I=1,2} \beta_I{\rm Tr~}U_{n_{\mu},\nu}(I)
U_{n_{\mu}+\hat{\nu},\rho}(I)
U_{n_{\mu}+\hat{\rho},\nu}^{\dagger}(I)
U_{n_{\mu},\rho}^{\dagger}(I)  \nonumber\\
&&+ C\sum_{n_{\mu},\nu} {\rm Tr~}X^{\dagger}(n_{\mu}+\hat{\nu})
U^{\dagger}_{n_{\mu},\nu}(1)X(n_{\mu})U_{n_{\mu},\nu}(2) 
+ c.c \nonumber\\
&&+ \sum_{n_{\mu}\in FP{\rm s}}V \left( {\rm Tr~}
X(n_{\mu})(i\sigma_3)X(n_{\mu})^{\dagger}(-i\sigma_3) \right)~,
\end{eqnarray}
\noindent
where $c.c$ means the complex conjugation and $\beta_I$ and $C$ are  
coupling constants. The potential term $V(x)$ 
is an even function for the $Z_2$-projected Polyakov loop
$X(n_{\mu})$ from the stick theorem.  
%%%
%
%%%% 

The effective action (\ref{eq:Seff}) is invariant under the FP gauge 
and stick symmetries. The  effective action suggests that 
the variable $X(n_{\mu})$ can be a candidate for 
the Higgs, which can play a role of a matter field in the 
fundamental representation of the FP gauge
symmetry. Since $X(n_{\mu})$ belongs to the fundamental representation,  
the confinement phase is expected to be connected with a Higgs
phase continuously from the Fradkin-Shenker's 
discussion\cite{Fradkin-Shenker}. Contrary to the usual continuum 
theory, the effective theory has two sets of four-dimensional 
gauge fields. The gauge fields on the $FP(1)$ and $FP(2)$ interact with  
each other by mediating $X(n_{\mu})$. After solving the mixing, we may
find a set of the four-dimensional gauge fields  
and of four-dimensional massive vector fields in the effective theory.

\section{Summary and Discussions}

In this paper, we have found a new symmetry (stick symmetry) on 
lattice gauge theory with $Z_2$-orbifolding. The 
symmetry and the associated theorem (stick theorem)
control the behavior of an order parameter
($Z_2$-projected Polyakov loop) and restrict the form 
of the effective action. It is found that the operator $X(n_{\mu})$ 
behaves like the Higgs field in the effective action. 
The definition of a Higgs field on lattice is an important problem. 
The field should be the fundamental representation of the FP gauge
symmetry. Although one of some candidates is $X(n_{\mu})$, better
candidates should be determined by requirements: the simpler form
and the smoothness for the continuum limit in calculating 
physical quantities such as Higgs mass.

When we consider an $SU(2)$ as the bulk gauge group, the stick symmetry 
belongs to a center in the $SU(2)$.  One may 
wonder whether the stick symmetry is always the same as the 
center of a bulk gauge group or not. The answer is clearly no. 
When $SU(3)$ is considered as the bulk gauge symmetry, we need 
to adopt $g_0={\rm diag}(1,-1,-1)$ by $Z_2$-orbifolding, by which 
the bulk gauge symmetry is broken down to $SU(2)\times U(1)$ at
the fixed points. In this case, we must generalize  
the new symmetry construction. We set a bulk gauge symmetry 
$G$. By $g_0\in G$, $G$ breaks down to a subgroup $H$ on the $FP(I)$  
\begin{equation}\label{eq:def-H}
H \equiv\{g\in G |~  g_0gg_0^{\dagger}=g \}~.
\end{equation}
\noindent
The {\it normalizer} $N_G(H)$ is defined as 
\begin{equation}\label{eq:normalizer}
N_G(H) \equiv \{ g\in G | ~ghg^{\dagger} =h'\in H ~
{\rm for}~^\forall h \in H \}~,
\end{equation}
\noindent
and $H$ is a normal subgroup of $N_G(H)$. It is clear that  
the stick transformation (\ref{eq:globalnewsymmetry}) is 
an element of $N_G(H)$ not $H$. More precisely, new symmetry 
up to the FP gauge symmetry is an element of the residual 
group $N_G(H)/H$. This residual group is discrete 
because $N_G(H)$ is isomorphic to $H$ as Lie groups. 
The $SU(3)$ case indicates that the residual group $N_G(H)/H$ 
is trivial not the center $Z_3$ of $SU(3)$. The change of the symmetry by 
$SU(3)$ has a serious influence on the role of the $Z_2$-projected Polyakov loop 
as an order parameter.   
%%%%%%%%%%%%%%
% 
%%%%%%%%%%%
With the $SU(N)$ bulk gauge symmetry, we find VEV of the loop, 
\begin{equation}\label{eq:SU(N)PLstronglimit}
<L_2> \rightarrow    \frac{|{\rm~Tr~g_0}|^2}{N}~, 
\end{equation}
\noindent
in the strong coupling limit. Since ${\rm Tr~g_0}$ is $-1$  
in the $SU(3)$ case, $<L_2> $ is non-vanishing in the 
limit. From this fact, it is difficult to treat $L_2(n_{\mu})$ as 
the order parameter for the stick symmetry generally. For general bulk 
gauge groups, the construction of its new symmetry is an open question.     

In the relation of our lattice model to 
the continuum theory, we have to make a few comments. 
In this article, we have considered the lattice model 
as a cutoff theory following to Irges and 
Knechtli\cite{Irges-Knechtli-1,Irges-Knechtli-2,Irges-Knechtli-3}. 
Although it is generally difficult to take 
the continuum limit of  lattice models, 
the realization of the limit is expected by the 2nd order phase transition. 
The further  analysis of the phase structure of 
the $S^1/Z_2$-orbifolded gauge theory may open its possibility.

\section*{Acknowledgements}
%
%We would like to thank ...........
%
The numerical calculations were carried out on SX8 at YITP in Kyoto 
University and on a supercomputer at Research Center for Nuclear 
Physics in Osaka University.
This work is supported in part by the Grant-in-Aid 
for Scientific Research (No.20540274(H.S.), No.21540285(K.T.)) by
the Japanese Ministry of Education, Science, Sports and Culture.

%\appendix
%\section{First Appendix} %Empty argument \section{} yields `Appendix'. 
%
%\section{Second Appendix}

\end{document}